# Nanophotonic engineering of far-field thermal emitters

Denis G. Baranov,[1,§] Yuzhe Xiao,[2,§] Igor A. Nechepurenko,[3] Alex Krasnok,[4] Andrea Alù,[4,5,6] and Mikhail A. Kats[2,7,8*]

[1]*Department of Physics, Chalmers University of Technology, 412 96 Gothenburg, Sweden*

[2]*Department of Electrical and Computer Engineering, The University of Wisconsin-Madison, Wisconsin, USA*

[3]*Dukhov Research Institute of Automatics, 22 Sushchevskaya, Moscow 127055, Russia*

[4]*Photonics Initiative, Advanced Science Research Center, City University of New York, New York 10031, USA*

[5]*Physics Program, Graduate Center, City University of New York, New York 10016, USA*

[6]*Department of Electrical Engineering, City College of The City University of New York, New York 10031, USA*

[7]*Department of Materials Science and Engineering, The University of Wisconsin-Madison, Wisconsin, USA*

[8]*Department of Physics, The University of Wisconsin-Madison, Wisconsin, USA*

## Abstract

Thermal emission is a ubiquitous and fundamental process by which all objects at non-zero temperatures radiate electromagnetic energy. This process is often presented to be incoherent in both space and time, resulting in broadband, omnidirectional light emission toward the far field, with a spectral density related to the emitter temperature by Planck's law. Over the past two decades, there has been considerable progress in engineering the spectrum, directionality, polarization, and temporal response of thermally emitted light using nanostructured materials. This review summarizes the basic physics of thermal emission, lays out various nanophotonic approaches to engineer thermal-emission in the far field, and highlights several relevant applications, including energy harvesting, lighting, and radiative cooling.

# I: Introduction

Every hot object emits electromagnetic radiation according to the fundamental principles of statistical mechanics. Examples of this phenomenon—dubbed thermal emission (TE)—include sunlight and the glow of an electric stovetop or embers in a fire. The basic physics behind TE from hot objects has been well understood for over a century, as described by Planck's law[1], which states that an *ideal black body* (a fictitious object that perfectly absorbs over the entire electromagnetic spectrum and for all incident angles) emits a broad spectrum of electromagnetic radiation determined by its temperature. Increasing the temperature of a black body results in an increase in emitted intensity, and a skew of the spectral distribution toward shorter wavelengths. A fundamental property of this process is that the thermal emissivity of any object—which quantifies the propensity of that object to generate TE compared to a black body—is determined by its optical absorptivity[1]. The connection between absorption and thermal emission, linked to the time reversibility of microscopic processes, suggests that the temporal and spatial coherence of the TE can be engineered via judicious material selection and patterning.

Engineering of TE is of great interest for applications in lighting, thermoregulation, energy harvesting, tagging, and imaging. This review summarizes the basic physics of TE and surveys recent advances in the far-field control of TE via nanophotonic engineering. First, we present the basic formalism that describes TE, and review realizations of narrowband, directional, and dynamically reconfigurable far-field thermal emitters based on nanophotonic structures. Then we review applications of engineered far-field TE, with emphasis on energy and sustainability.

# II: Theoretical background

At elevated temperatures, the constituents of matter, including electrons and atomic nuclei, possess kinetic energy. The motion of these charges is continuously changing due to various particle-particle interactions (e.g., inter-atomic collisions), resulting in fluctuating currents. The electromagnetic energy radiated by these fluctuating currents is referred to as TE. The precise nature of TE, including its intensity, spectral and angular distributions, and polarization, depends not only on the temperature of the emitter, but also on its optical properties.

TE of an optically large object can be characterized by its spectral radiance $L^l(\theta, \phi, \nu, T)$, defined as the emitted optical power per unit area, per unit solid angle in the direction $(\theta, \phi)$, per unit frequency at frequency $\nu$, in polarization state $l$. A simple expression for the spectral radiance can be derived for a black body, which can be approximated in a laboratory by a large enclosure with absorbing walls and a small hole through which light can enter and exit, or by highly absorbing optical structures that suppress reflection and scattering, such as vertically oriented carbon nanotube arrays[2]. The expression for the spectral radiance of a black body, $L_{BB}(\nu, T)$, is known as Planck's law[1]:

$$L_{BB}(\nu, T) = \frac{c}{4\pi} \frac{8\pi\nu^2}{c^3} \frac{h\nu}{e^{\frac{h\nu}{k_B T}} - 1}, \tag{1}$$

where $c$ is the speed of light, $h$ is Planck's constant, and $k_B$ is the Boltzmann constant. The spectral radiance of a black body is independent of angle and polarization state.

Equation 1 can be understood by considering a black body in thermal equilibrium with a surrounding photon reservoir in free space [Fig. 1(a)], such that the emitted energy density matches the energy density of the reservoir[3]. In free space, the density of states for photons with energy $h\nu$ is $8\pi\nu^2/c^3$. Meanwhile, according to Bose-Einstein statistics, the expected number of photons in a specific electromagnetic mode at frequency $\nu$ is $\left(e^{\frac{h\nu}{k_B T}} - 1\right)^{-1}$. The factor $c/4\pi$ relates energy density and radiance. Note that Eq. 1 describes black-body emission in three-dimensional free space, and it can be generalized for emission into a material with a given refractive index[3], or into a space with an arbitrary number of dimensions[4]. For example, the one-dimensional version of Planck's law describes the Johnson-Nyquist voltage noise generated across a resistor at a finite temperature[5]. The close connection between thermal emission and Johnson-Nyquist noise can be visualized by considering the energy balance between radiation incident on (and emitted by) an antenna connected to a resistor at a finite temperature[6].

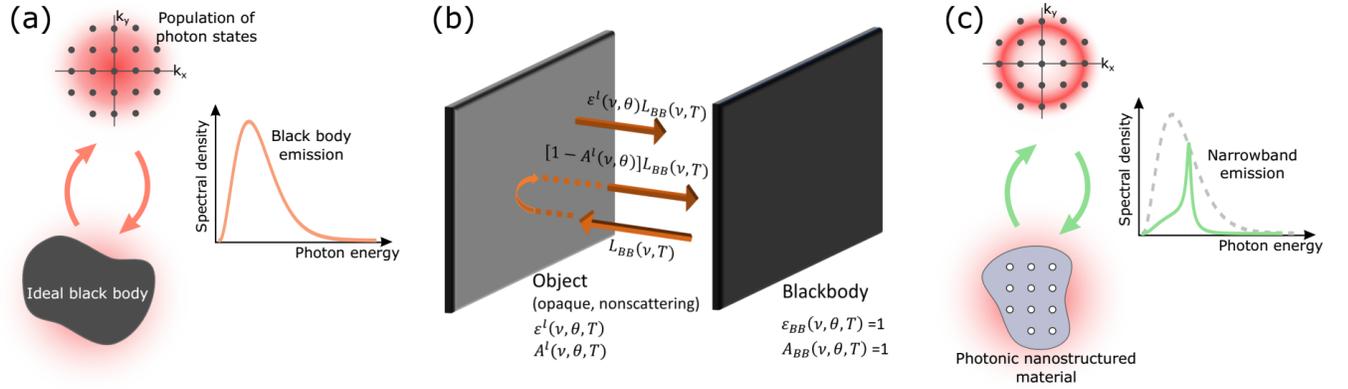

**Figure 1. Physics of thermal emission (TE).** (a) An ideal black body establishes thermal equilibrium with a photon bath by absorbing and radiating photons. Thermal photons occupy the available far-field states according to the Bose-Einstein distribution, resulting in the Planck spectrum. The points depict selected photon states in momentum space, reduced to two dimensions to simplify visualization. (b) Schematic of the energy balance described by Kirchhoff's law. As an example, an opaque non-scattering object (i.e., absorption *A* and reflectance *R* add up to 1) is in thermal equilibrium with a black body at temperature *T*. According to the second law of thermodynamics, the net energy flow between them must be zero, and therefore the emissivity $\varepsilon = A$. (c) TE from a nanostructured material with frequency-dependent absorptivity can result in the narrowing of the spectrum.

Integrating Eq. 1 over all frequencies and angles results in the Stefan-Boltzmann law, which relates the total irradiance *I* of a black body to its temperature[7]:

$$I = \left(\frac{2\pi^5 k_B^4}{15 c^2 h^3}\right) T^4 \qquad (2)$$

When emission occurs into a transparent dielectric medium, the total emitted power is proportional to the medium permittivity[3] because of the modified density of photonic states in the environment. A similar integration of the one-dimensional "Planck law" that describes Johnson-Nyquist noise yields the one-dimensional "Stefan-Boltzmann law", $P = \left(\frac{\pi k_B^2}{3\hbar}\right) T^2$, that describes the total power generated by thermal noise in a short-circuited resistor [6].

The TE spectrum of any realistic object deviates from that of a black body. Various techniques can be used to calculate TE in both near- and far-field regimes from emitters with arbitrary temperature distributions and optical responses[8,9]. In the far field and for a well-defined temperature $T$, the calculations can be simplified by using the concept of emissivity, $\varepsilon^l(\nu, \theta, \phi, T)$, defined as the ratio of spectral radiance of TE from an object to that of a black body:

$$\varepsilon^l(\nu, \theta, \phi, T) = \frac{L^l(\nu,\theta,\phi,T)}{L_{BB}(\nu,T)} \tag{3}$$

For a black body, $\varepsilon^l(\nu, \theta, \phi, T) = A^l(\nu, \theta, \phi, T) = 1$, where $A^l(\nu, \theta, \phi, T)$ is the absorptivity. For an arbitrary reciprocal object, the emissivity and absorptivity are equivalent, as described by Kirchhoff's law of thermal radiation[10]:

$$\varepsilon^l(\nu, \theta, \phi, T) = A^l(\nu, \theta, \phi, T) \tag{4}$$

Kirchhoff's law captures the direct relationship between optical absorption and TE, and it can be understood via the simple thought experiment shown in Fig. 1(b): two objects at the same temperature $T$ are placed next to each other—one black body, and one defined by some absorptivity $A$ and emissivity $\varepsilon$. Here, we assume that the objects are both opaque, non-scattering, and surrounded by a low-temperature vacuum, though these assumptions can be relaxed without affecting the conclusions. Because the two objects are at the same temperature, there can be no net energy flow between them, resulting in a precise balance of emission and absorption. Balancing the emitted, reflected, and absorbed components yields Kirchhoff's law.

The direct relation between absorptivity and emissivity of an object (Eq. 4) is the key result that enables engineering of TE using nanophotonic structures (Fig. 1c). In the following sections, we summarize various ways of achieving narrowband, directive and/or polarized TE, describe approaches to enable dynamic tunability, and review applications that benefit from such extensive control of TE.

**Temperature.** To understand certain subtleties of TE, it is important to consider the precise meaning of the temperature of an emitter. In the simplest case, all parts of an emitting object are at equilibrium, and the temperature is well defined; the Planck and Kirchhoff laws (Eqs. 2-3) can then be trivially used. Instead, a macroscopic emitter may have a position-dependent temperature distribution throughout its volume; in this case, the various parts of this object are not in thermal equilibrium, but the total emitter can be described as an assembly of smaller emitters, each with a roughly uniform temperature. Kirchhoff's law can be generalized to include such position-dependent temperature distributions[11,12], or the total TE may be calculated directly using the fluctuation–dissipation theorem[8,9]. Furthermore, an emitter may comprise one or more spatial regions not in internal equilibrium; for example, the temperature of the carriers in a semiconductor or metal might be different from the lattice temperature[13], and in extreme cases the temperature may not be well defined due to non-thermalized carrier distributions[13]. Thus, there exists a continuum relating what is commonly referred to as TE, i.e., emission from an object in equilibrium, to other types of light emission, e.g., hot-carrier luminescence[14].

## III: ENGINEERING THERMAL EMISSION

**Box 1—Coherence of thermal emission**

Coherence is a fundamental property of wave processes. Broadly, the degree of coherence determines the ability of a wave to generate interference.

*Temporal coherence* describes the degree to which a light wave interferes with a time-delayed version of itself at some point in space. The temporal coherence of a classical wave process can be characterized by the first-order correlation function $g(\tau) = \langle E^*(t)E(t+\tau)\rangle/\langle|E(t)|^2\rangle$, where $\langle...\rangle$ indicates ensemble/statistical averaging[15]. The coherence time $\tau_c$ (or, correspondingly, the temporal coherence length $L_c = c\tau_c$) is defined as the time after which $g(\tau)$ decays by a factor of $1/e$, and it is roughly equal to the time span over which a wave can be approximated as sinusoidal (i.e., with a particular frequency). Black-body TE has a wide spectrum, and conversely a short coherence length compared to its mean wavelength[16]. In contrast, a narrowband emission spectrum implies a longer coherence time. However, the enhanced coherence time/length of narrowband TE comes at the price of a reduced integrated intensity, since Planck's law limits the spectral energy density at each wavelength. This is in contrast to the behavior of lasers, where emission-line narrowing often corresponds to increasing the overall output[15].

*Spatial coherence* expresses the correlation of fields at two different positions at the same instant in time, describing the ability of a light wave to interfere with a spatially-shifted version of itself. Like temporal coherence, a correlation function can be used to characterize the spatial coherence: $g(\mathbf{r}, \mathbf{r}+d\mathbf{r}) = \langle E^*(\mathbf{r})E(\mathbf{r}+d\mathbf{r})\rangle/\sqrt{I(\mathbf{r})I(\mathbf{r}+d\mathbf{r})}$, where $I$ is the intensity[17]. Spatial coherence can be quantified by a *spatial* coherence length $L_{sc}$, which is the distance around a point $\mathbf{r}$ over which the spatial correlation function $g(\mathbf{r}, \mathbf{r}+d\mathbf{r})$ decays by a factor of $1/e$. For an emitter with very weak or no spatial coherence, the TE will be isotropic. Instead, pronounced spatial coherence of a thermal emitter can lead to highly directive TE which can, e.g., be observed in experiments with gratings[18–20], as described below. For many sources, the spatial coherence length $L_{sc}$ can be estimated from the far-field emission angle $\theta$ at wavelength $\lambda$ through $L_{sc} \sim \lambda/\theta$ [15,18]. Lastly, polarization is also related to coherence: fields in orthogonal

> polarization states cannot interfere.

**Narrowband, directive and polarized far-field emission.** A variety of applications require control over the TE spectrum, directionality, and polarization, and many modern tools of optics and photonics have been brought to bear on this engineering challenge.

Perhaps the simplest way to achieve narrowband TE (i.e., TE with high temporal coherence) is to design an emitter comprising a bulk material that has narrowband optically active resonances with large oscillator strengths, resulting in narrowband absorptivity/emissivity peaks. Examples of such bulk materials include, but are not limited to, polar dielectrics (such as SiC) in the mid infrared[21] and rare-earth oxides (such as erbium oxide and ytterbium oxide) in the visible to near-infrared range[22]. Quantum confinement in low-dimensional materials can similarly be used to achieve narrow resonances; examples include individual (or highly homogeneous) carbon nanotubes (CNTs), which can radiate narrowband TE from the visible to the infrared due to interband transitions[23], multiple quantum wells with narrow interband transitions[24], and atomically thin semiconductors such as $MoS_2$ [25]. On the other hand, if a selective emitter with a step-like emissivity is desired, a wider range of simple bulk materials is available, including composite materials such as cermets, and metals above the plasma frequency (see Ref. [[26]] for an overview).

A more flexible way to control TE is to use a combination of material and structural optical resonances to enhance and/or suppress emission. For example, optical interference in few-layer lossy thin films allows for control over the position and amplitude of absorption peaks [27] and thus TE, potentially with a narrow bandwidth[28,29]. Photonic crystals can also offer substantial control of the TE bandwidth. For example, a truncated one-dimensional[30] or three-dimensional[31] photonic crystal may be placed on top of an emitting layer, and serve as a passive filter that reflects TE at photonic-bandgap frequencies [Fig. 2(a)]. The emitting material can also be incorporated into the photonic crystal itself[32].

Alternatively, a truncated photonic crystal may be placed at some distance above the emitting substrate, forming a cavity, leading to enhanced emissivity at the resonant frequency [33,34].

> **Box 2—Critical coupling**
>
> According to Kirchhoff's law, enhancing the emissivity of a structure is equivalent to enhancing its absorptivity and minimizing reflection, transmission, and scattering—a classic problem in optics and electromagnetics. A particularly helpful tool in this context is the concept of critical coupling[35].
>
> Any linear electromagnetic system can be described by a set of modes, each characterized by radiative $\gamma_{rad}$ and non-radiative $\gamma_{nr}$ decay rates. The former is the rate at which energy from the mode is radiated into the surroundings, whereas the latter is the rate at which energy is absorbed. In a single-mode system, perfect absorption arises at the resonant frequency under the critical coupling condition, when the two decay rates are equal: $\gamma_{rad} = \gamma_{nr}$[35]. Correspondingly, such a system will exhibit unity emissivity at the resonant frequency with a linewidth of $2\gamma_{rad}$. This concept therefore provides a useful tool for TE engineering with resonant structures[33,36,37], and suggests a way toward strong narrowband emission[38]. For example, near-critical coupling to a photonic crystal consisting of an array of holes in a multiple-quantum-well slab was used to achieve high emissivity and quality factors of approximately 100, resulting in very narrowband TE [Fig. 2(b)].[24]

Arrays of resonant nanostructures with subwavelength separation—sometimes referred to as metamaterials or metasurfaces, depending on the context—have also been used to realize narrowband TE across the visible, infrared, and terahertz ranges. Examples include arrays of cross-shaped metallic nanoantennas[39] and metal stripes[40,41] on top of metallic ground planes separated by dielectric spacers, slits in metallic surfaces[42], and semiconductor nanorods[43] [Fig. 2(c)]. Narrowband TE in the terahertz region has also been demonstrated in a system that strongly couples mid-infrared interband transitions to a plasmonic microcavity, pushing a polariton branch to terahertz frequencies[44].

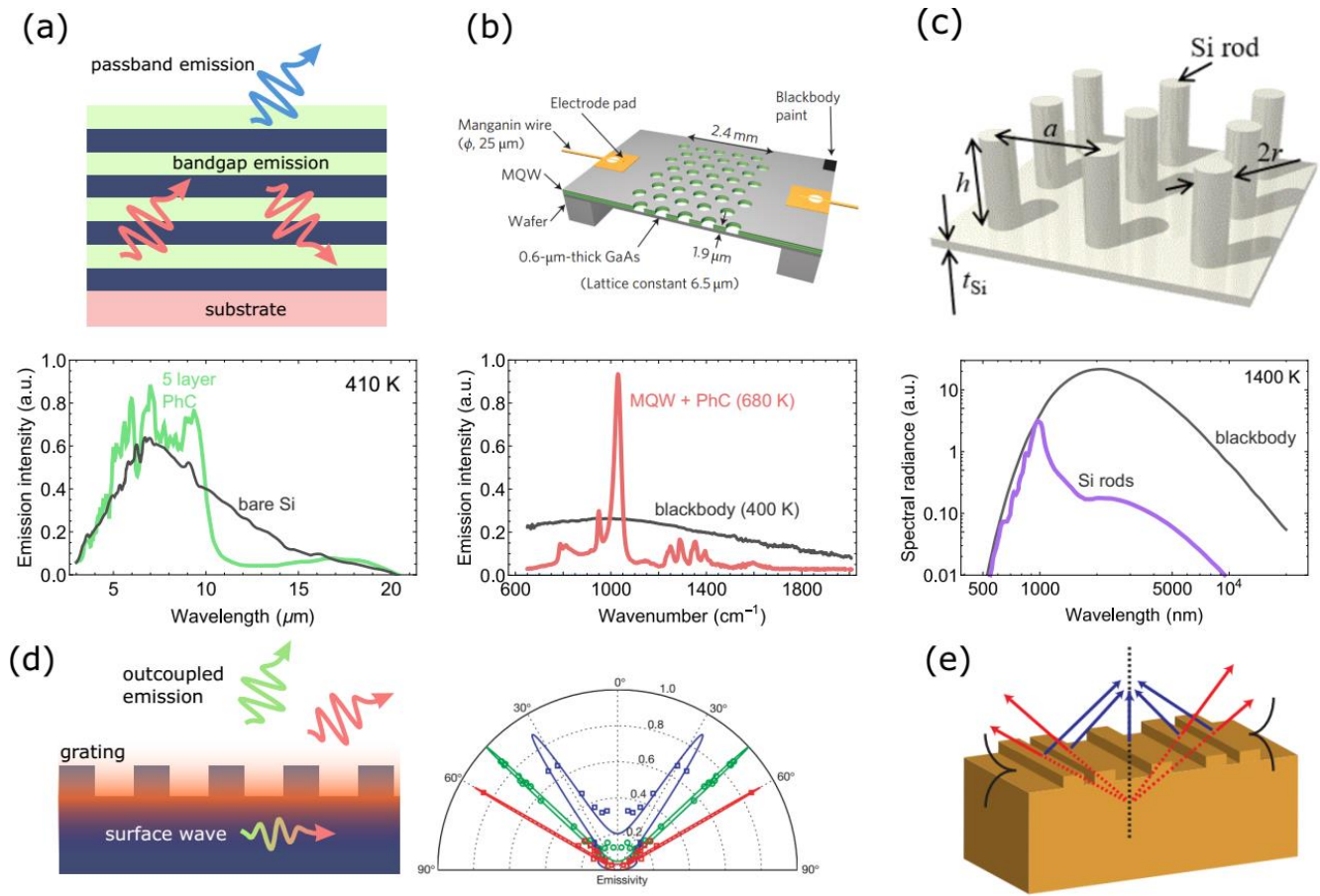

**Figure. 2. Narrowband and directive TE from nanophotonic systems**. (a) Frequency-selective TE enabled by a photonic-crystal filter, transmitting TE from the substrate within the propagation band and reflecting the light that falls within the bandgap. Bottom: measured TE spectrum from an emitting layer heated to 410 K covered with a 3D photonic crystal. (b) Top: a photonic-crystal slab etched out of a multiple-quantum-well layer, which results in narrowband TE due to the coupling of a narrowband material resonance and a high-Q photonic mode. Bottom: the resulting TE spectrum at 680 K compared to that of a black body at 400 K. (c) Top: an array of silicon rods enables selective TE by combining the intrinsic electronic resonance of the material and the photonic resonances of the rods. Bottom: the resulting spectral radiance of the Si rods array at 1400 K. (d) Left: directive TE enabled by the Rayleigh anomaly of a grating, which outcouples a thermally generated surface wave to free space. Right: angular distribution of TE from a SiC grating at three wavelengths close to the Rayleigh anomaly. (e)

Schematic of focused TE produced by a nanostructured SiC surface. The non-uniform distribution of scattering phase enables the focusing of light outcoupled from a thermally excited surface wave. Images reprinted with permissions from: a- ref. [31]; b - ref. [24]; c - ref. [43]; d - ref. [18] ; e - ref. [45].

The structures discussed above enable spectral-bandwidth engineering of TE. For many applications, directional control of TE also adds an important degree of freedom.

Directional TE was first observed from gratings patterned into doped silicon, resulting in selective out-coupling of surface plasmons on the doped-silicon surface into free space[46]. Under emittance angles satisfying the momentum matching condition, $k_0 \sin \theta = k_x + mG$ (where $k_0$ is the free-space wavenumber, $\theta$ is the emission angle, $k_x$ is the guided-mode wavenumber, and $G$ is the reciprocal lattice vector), coupling is enabled between a guided-mode resonance and free-space light (i.e., a Wood or Wood-Rayleigh anomaly[47]) [Fig. 2(d)]. Such directional emission has been demonstrated with the use of SiC[18], tungsten[19], and SiO$_2$[20] gratings in the infrared range. Due to the dependence on the emission angle of the free-space wavelength at which coupling is achieved, the system can thermally emit a "rainbow" with light at different wavelengths being radiated toward different directions. If the surface waves can only exist in a narrow frequency range, resulting in long-range surface-wave correlations[48], then this rainbow will have few frequency components, and thus relatively high temporal and spatial coherence can be achieved simultaneously[18]. The resulting TE exhibits enhanced directivity (and hence enhanced spatial coherence; see Box 1). In most demonstrations, frequency selectivity is achieved using narrowband material resonances, e.g., in SiC[18], and then angular selectivity is enforced by the structure. Alternatively, the frequency selectivity can be engineered using subwavelength resonators, and angular selectivity can then achieved via array effects[49]. A different strategy must be employed to achieve angularly selective but broadband TE, e.g., using non-resonant metamaterials with directional absorption/emissivity due to a Brewster-like response[50].

Control of the spatial coherence of TE can be generalized to shape the far field of thermal emission in more-complex ways. For example, by patterning an array of antennas with a non-uniform scattering phase distribution on a SiC surface, free-space focusing of TE has been demonstrated [Fig. 2(e)][45]. Guided by a similar principle, a non-uniform metasurface composed of thermal emitters with varying angular emissivity patterns was shown to suppress TE into a certain area, forming a "dark spot"[51]. A recent theoretical study demonstrated TE from objects comprising an epsilon-near-zero material, enabled by the enhanced spatial coherence resulting from the stretching of the wavelength inside the epsilon-near-zero material [52].

The polarization of TE can also be controlled using engineered emitters. In the case of linear polarization, this is relatively straightforward; in fact, most engineered thermal emitters are at least partially polarized (e.g., those of refs. [18,19,53]). More deliberately, anisotropic plasmonic arrays with high absorptivity for one polarization and low absorptivity for the orthogonal polarization have been employed to obtain polarized and narrowband emission[42,54]. A quarter-wave retarder can be added to such structures to achieve circularly-polarized TE[55]. Circular-polarization selectivity can also be directly achieved in a structured emitter; e.g., in resonant silicon metasurfaces with planar chirality [56].

We conclude this section by noting that there is, in principle, no theoretical limit on the spectral or angular bandwidth of thermal emitters. For example, the concept of bound states in the continuum can enable very narrowband absorption peaks with highly directional patterns [57] (see Box 2). In practice, however, the spectral and angular bandwidths of TE of deliberately narrowband emitters will be eventually limited by material loss and fabrication imperfections, as well as by the emitter size.

**Tunable thermal emission.** In the previous section, we reviewed techniques to engineer far-field TE in the spectral, angular, and polarization domains. Here, we consider techniques to modulate TE in time. Because TE depends on both emissivity and temperature (Eqs. 1 and 3), the modulation of either can be used to achieve dynamic control.

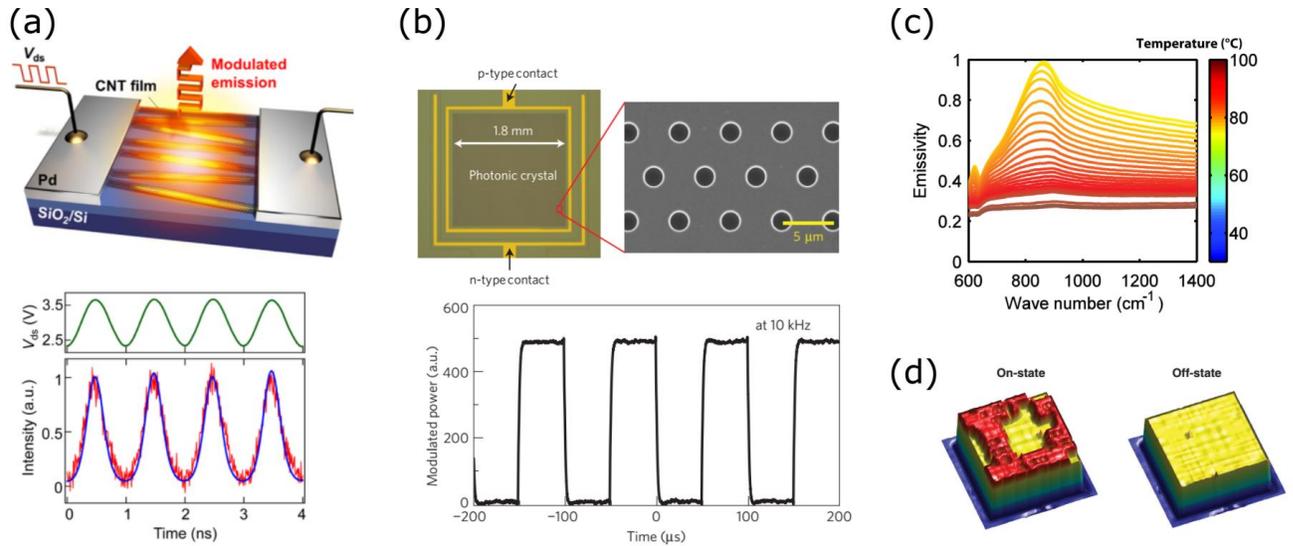

**Figure 3. Dynamic modulation of TE.** (a) Top: A fast electrically controlled thermal emitter based on the small heat capacity of a carbon nanotube (CNT) film and its high heat-dissipation rate to the substrate, resulting in fast temperature change. Bottom: Experimental emission modulation (red) under a continuous input voltage of 1 GHz (green). (b) Top: Dynamic thermal emitter based on a p-i-n GaAs diode incorporating a two-dimensional photonic-crystal slab and GaAs/Al$_{0.3}$Ga$_{0.7}$As quantum wells. Modulation of TE is achieved through emissivity, which is controlled by a gate voltage on the quantum wells. Bottom: Temporal waveforms of the TE power at a modulation frequency of 10 kHz. (c) Temperature-dependent emissivity of a thin film of VO$_2$ on sapphire, with the tunability enabled by the insulator-to-metal phase transition of VO$_2$. (d) Infrared images of a 64-pixel tunable-emitter array, comprising metamaterial microelectromechanical elements with two states, with the tuning via an electric bias between the ground plane and the top layer. Images reprinted with permissions from: a - ref. [58]; b - ref. [59]; c - ref. [60]; d - ref. [61].

The most straightforward way to modulate TE is via the temperature of the emitter. However, in many cases, modulation much faster than hundreds of Hz is difficult to realize, limited primarily by the large thermal time constants in macroscopic structures[62]. The speed can be increased by decreasing the volume, and hence the heat capacity, of the emitters; however, this scaling also frequently decreases the total emitted power, which is proportional to the emitting area. This trade-off can be circumvented using emitters with inherently small volumes but relatively strong interaction with the incident light, such as two-dimensional materials or nanoantennas[63]. For example, current injection into carbon-nanotube (CNT) films has been used to demonstrate TE modulation of up to 10 GHz [Fig. 3(a)] [58]. The integration of such emitters into nanophotonic cavities has been used to realize narrowband light sources operating at GHz speeds[64].

A different way to increase the temperature modulation speed is to forgo heating the lattice of the material comprising the emitter, and thus dramatically reduce the heat capacity. For example, when pumped by a laser, electrons can be driven far out of thermal equilibrium with phonons, resulting in an electron temperature much higher than the lattice temperature[65]. Depending on the material, these temperatures can reach tens of thousands of degrees without resulting in damage[66], greatly enhancing the thermally emitted power. TE modulation from hot electrons has been demonstrated using ultrafast pump pulses, with modulation speeds in the 100s of femtoseconds in graphene[66], and several picoseconds in metals such as gold and tungsten[67].

Dynamic control can also be achieved without temperature modulation via a time-dependent emissivity. The emissivity can be made tunable by using materials that can be tuned via the application of voltage (i.e., electrochromics[68]), heat (thermochromics[69]), strain[70], etc. Early demonstrations of tunable TE used electrochromic materials such as tungsten trioxide and nickel oxide[71]. More recently, electrical control of emissivity has been demonstrated using carrier-density tuning in a gated quantum well in proximity to a grating[72]. A similar design combining a high-quality-factor photonic-crystal slab with a gated multiple-quantum-well structure resulted in 600 kHz modulation of narrowband TE [Fig. 3(b)] [59]. This concept was extended to resonant graphene antennas, where the plasmonic response of the graphene was modulated via a gate[73], and to doped semiconductors[74].

One may also employ all-optical modulation of the emissivity via photocarrier doping. For example, optical pumping of silicon has been used to achieve pulsed TE, with the speed limited by the ~200 $\mu$s free-carrier lifetime[75]. A recent work demonstrated nanosecond modulation of emissivity of a gallium-arsenide (GaAs) surface, enabled by the much-shorter free-carrier lifetime in GaAs compared to silicon[76]. A similar approach was also used to realize simultaneous spatial and temporal control of TE via photodoping of zinc oxide, but with a much slower speed[77].

The emissivity can also be modulated using thermochromics—materials that undergo large reversible changes in their optical properties as a function of temperature. Recently, significant efforts have been focused on the use of germanium-antimony-tellurium (GST), which can exist in amorphous and crystalline phases, and can be switched back and forth via fast and slow anneals[78]. GST has been integrated into structures incorporating plasmonic antennas[79] and thin films[78], in each case resulting in a large change in emissivity corresponding to switching of the GST phase. Similarly, materials with electronic phase transitions can be used for emissivity control. For example, the simultaneous structural and electronic phase transition in vanadium dioxide ($VO_2$) results in a large change of optical constants across the transition[80], and has been used to realize structures with temperature-tunable emissivity, including some featuring negative-differential TE (i.e., the emission decreases with increasing temperature) [Fig. 3(c)][60]. More recently, the more-gradual but nevertheless large electronic phase transition in samarium nickelate ($SmNiO_3$) has also been used for TE management[81].

Mechanical strain can similarly be used to modulate the emissivity[82]. In one realization, an emitter with temperature-dependent emissivity was engineered using a patterned metallic structure suspended over a back reflector, such that the gap size depended on temperature via thermal expansion[83]. Strain can also be actuated using electrostatic forces; for example, modulation of TE was recently demonstrated via an electric bias between the ground plane and the top layer of a nano-structured thermal emitter [Fig. 3(d)][61]. Lastly, emissivity can be controlled via an applied magnetic field; for example, magnetic-field-induced optical anisotropy of a semiconductor has been used to modify the spectrum of TE[84].

In general, the limit of the modulation rate of TE is determined by two aspects: (i) how quickly the stimulus can be applied to the emitter, and (ii) how quickly the emitter responds to the stimulus. The first factor reflects the speed of the external control device; in the literature, the fastest modulation has been enabled by optical pumping with ultrafast pulses on the scale on hundreds of femtoseconds[66]. In the case of emissivity modulation, the second factor is determined by the material response, which could be nearly instantaneous (*e.g.,* electro-optic or magneto-optic modulation).

**Emission from subwavelength structures.** Though Eqs. (1) and (3) describe far-field TE into free-space from an arbitrary emitter, extra care must be taken when the transverse dimensions of the emitter become comparable to or much smaller than the emission wavelength.

The TE from an optically small object depends on its absorption cross section.[85] Since objects smaller or comparable to the wavelength often have absorption cross sections much larger than their geometric cross sections, TE *per geometric area* in the far field may appear to be 'super-Planckian'[85]. There are two equivalent ways to reconcile this phenomenon with Planck's and Kirchhoff's laws. The first option is to assume that the area considered is the geometric area; in this case, the emissivity can be allowed to exceed one, and can even become arbitrarily large, since a large absorption cross section can be achieved via proper engineering, even when the physical cross section is small[85,86]. For example, the connection between emissivity and the absorption cross section was experimentally demonstrated using SiC nanorods with resonances in the mid infrared[53]. The second option is to consider the effective area to be the absorption cross section, rather than the geometric area. In this case, the emissivity of a thermal emitter cannot exceed one.

**Box 3—Near-field thermal emission and its extraction**

For TE in the near-field regions of emitters, e.g., close to interfaces supporting guided waves, Planck's law (Eqs. 1 and 3) does not provide an adequate description, because it is limited to light accessing the far-field radiation modes. For example, near-field TE can transfer substantially more energy across subwavelength gaps via evanescent fields than what would be expected from the Planck black-body limit[48,87]. This effect, referred to as near-field heat transfer, has been observed in a variety of systems with nanoscale gaps[88,89] and has led to a number of intriguing applied proposals, such as thermal diodes, transistors, and memory elements. Near-field TE is not the focus of this review, and extensive overviews of the field can be found in, e.g., ref. [[90]].

Various schemes can be used to outcouple near-field TE to propagating fields. For example, a nanoparticle or a microscope tip placed close to a surface can scatter evanescent fields into the far field [91,92] (Figure below, left). A similar strategy can be used to enhance the extraction of TE from a region of space with a large density of states using a non-absorbing high-index dome (Figure below, right) [93]; more TE is emitted into the high-index material, which can then escape into free space due to the curvature of the dome. Note, however, that that the power outcoupled using any TE extraction scheme is ultimately bounded by Planck's law, considering the interaction area of the entire structure comprising the emitter and the extractor.

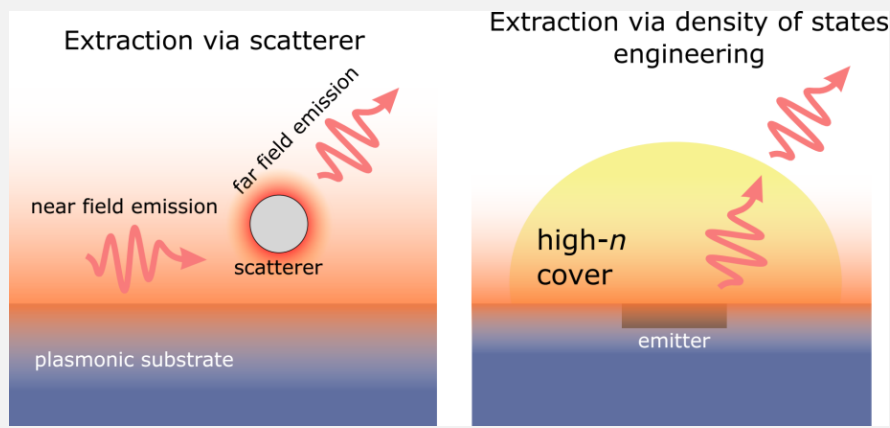

## Applications

**Energy harvesting and lighting.** Some of the most technologically significant applications of engineered TE are in the field of energy conversion, including energy harvesting and lighting.

Thermophotovoltaic (TPV) technology uses photovoltaics (PVs) to convert engineered TE from hot objects into electricity, and is being explored for applications that include direct solar-energy conversion [94–96], storage and retrieval of solar-thermal energy [97], recycling of waste heat from power plants and combustion engines [98], combined heat-and-power generation [99], and harvesting of TE from the Earth [100]. Due to space constraints, in this review we focus on direct solar-energy harvesting with TPVs, and elaborate on both the potential advantages and technical challenges of TPV for this application.

In solar-energy harvesting, the conventional PV technology has several well-known drawbacks that limit the efficiency to far below the ~95% Carnot efficiency limit (calculated given the temperature differential between the sun and the earth) [94]. In single-junction PVs, only above-bandgap photons are absorbed; furthermore, the difference between the photon energy and the band gap is lost to heat. These loss mechanisms, together with additional losses due to radiative recombination, cap the theoretical efficiency of single-junction PVs for non-concentrated sunlight at ~33% (similarly, two-junction cells are capped at ~42%, and three-junction at ~49%); this cap is known as the Shockley-Queisser limit [94]. These limits can—in principle—be overcome by using solar TPVs. In TPVs, solar photons are first absorbed by an intermediate broadband absorber, raising its temperature, and the energy is then re-directed toward a conventional PV using a frequency-selective thermal emitter [Fig. 4(a)].[26]

Assuming highly idealized components—ideal geometry, TE engineered to be monochromatic TE, only radiative recombination in the PV cell, full solar concentration, etc.—the efficiency of a TPV system could be as high as 85% [101]. In more realistic geometries, many factors affect the efficiency of a solar TPV device, including the absorber/emitter spectral responses and the ratio of their areas, parasitic thermal loss, nonidealities in the PV cell, among others. In particular, the temperature of the intermediate absorber/emitter is an important parameter in determining the efficiency of solar TPV devices [101]. TE engineering, therefore, plays a key role on both ends of solar TPV systems—the solar absorbers and the selective emitters [26]. The ideal solar absorber should capture the vast majority of the solar spectrum—in the ultraviolet, visible, and near infrared—but should suppress TE in the mid infrared, such that energy loss to the sky is minimized. Further improvement of the solar absorber is possible by limiting the angular range of absorption/emission to the angular range of sunlight, or using concentrated sunlight [101]. At the same time, the ideal emitter should have a relatively narrow emissivity peak slightly above the PV bandgap [Fig. 4(b)], such that maximal conversion efficiency is achieved at the PV element [95].

A number of studies have employed nanophotonic engineering in the design of solar TPV systems [95,96,102,103]. For practical area ratios between the emitter and the absorber (~20), solar concentrations (~1000), and for realistic nanophotonic emitters, it has been numerically demonstrated that efficiencies of ~40% can be achieved using a single-junction PV, overcoming the Shockley-Queisser limit [95]. These numbers, however, have not yet been approached in the experiments, with the record TPV efficiency reaching 7% in systems incorporating a Si/SiO$_2$ photonic crystal selective emitter and a narrowband optical filter [96,103]. These numbers are far from the theoretical limit due to a combination of factors, including suboptimal spectral responses of the absorber and emitter, low efficiency of the PV cell (a bandgap of 0.55 eV was used to match the TE spectrum from the emitter, and these PVs are less mature compared to conventional PVs), and various parasitic loss mechanisms such as thermal conduction. For example, it was predicted that an overall efficiency of 20% can be realized by scaling up the device, which reduces the parasitic losses [103].

Nevertheless, given that a reasonably optimized theoretical TPV design yields system efficiencies of ~40% [95], it is not clear if the complexity of the TPV arrangement provides sufficient benefit over conventional multi-junction PV, which can achieve similar or greater efficiencies [94]. However, applications of TPVs in which the heat source is *not* the sun may be more promising, since existing technologies to convert, e.g., waste heat, are far less efficient than state-of-the-art PVs for solar-energy conversion [97–99]. We note that although near-field TE is largely outside of the scope of this review (Box 3), positioning the emitter and PV at subwavelength distances and engineering near-field TE may result in improved TPV performance, and reduce the temperatures required for high power densities [104].

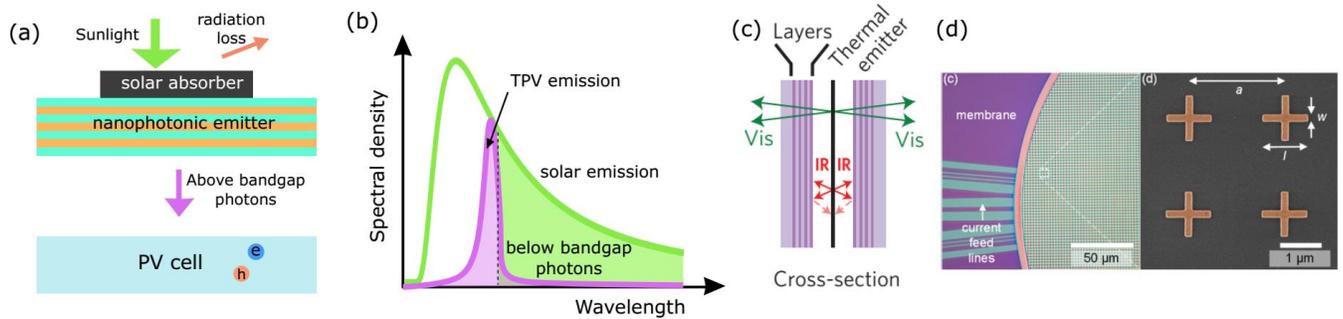

**Figure 4. Thermal emission control for energy conversion and lighting.** (a) Schematic of a solar thermophotovoltaic (TPV) device consisting of a broadband solar absorber, a selective thermal emitter (e.g., a one-dimensional photonic crystal), and a PV cell. For non-solar applications, a different heat source (e.g., waste heat from a power plant) replaces sunlight. (b) A large fraction of solar energy consists of photons with energies below the typical photovoltaic (PV) bandgap (1.1 eV for silicon) is not absorbed by the PV cell. In a solar TPV device, solar energy is converted to photons right above the PV bandgap via selective thermal emitter. (c) Nanophotonic coating for efficient incandescence: a thin-film multilayer transmits visible TE emitted by a filament, while reflecting in the infrared, leading to enhanced luminous efficiency. (d) A microscope image of a wavelength-selective infrared thermal emitter for $CO_2$ sensing. Images reprinted with permissions from: c – ref. [105]; d - ref. [106].

Nanophotonic structures can also be employed to increase the efficiency of lighting devices. In traditional incandescent lighting, which has gone out of favor for commercial lighting applications in parts of the world, a tungsten filament is heated up to 2000-3000 K, resulting in the majority of energy being emitted in the infrared part of spectrum rather than the visible. One approach to improving the luminous efficiency (visibility of a light source by a human eye relative to its power consumption) of an incandescent light is to decrease the infrared emissivity of the filament[107]. This can also be accomplished by decreasing the *effective* emissivity of the incandescent source by enclosing a high temperature filament/emitter with a filter that is reflective in the infrared but transparent in the visible [108]. Recently, luminous efficiencies of up to 40% have been demonstrated using this approach, which exceeds that of state-of-the-art LED lamps [Fig. 4(c)] [105].

TE engineering can also be used to design selective and relatively bright light sources in the mid-infrared portion of the spectrum. For example, narrowband TE produced by critically coupled selective emitters has been used as a light source for infrared gas sensing using, e.g., a cross-shaped antenna array [Fig. 4(d)] [106] or a thin platinum membrane placed $\lambda/4$ above a gold mirror [109].

The engineering challenges of both TPVs and lighting devices require advanced nanophotonic engineering, with the structures ultimately requiring high-temperature stability. Therefore, the materials comprising these emitters should have high melting points in both their bulk and nanostructured forms [110], and should not easily react with ambient air. Furthermore, in thermal emitters comprising multiple materials, the thermal expansion coefficients should be compatible [111]. Various refractory metals and dielectrics have been used for high-temperature TE engineering. For example, silicon carbide (SiC) gratings have been used to achieve directional TE [18]. The plasmonic properties of refractory metals are often used to design thermal emitters; for example, tungsten [112] and tantalum [113] photonic crystals, as well as tungsten hyperbolic metamaterials [114] have been designed as selective emitters for TPV applications. Selective emission has also been realized by doping refractory ceramic materials, such as zirconia and yttrium, with rare-earth dopants such as erbium, ytterbium, and others [115], which have narrowband absorption bands in the near infrared.

**Radiative cooling and textile heat management.** An appealing application of selective TE is the enhancement of radiative cooling—the process by which objects cool down when radiating light. This process requires a cold sink in radiative "contact" with the emitter, and the largest cold sink available is the universe, whose temperature is estimated to be around 3 K based on the measurements of the cosmic background radiation [116]. To maximize the rate of cooling via TE toward the cold universe, a radiative-cooling structure should exhibit maximal emissivity in the spectral range of 8 - 13 $\mu$m, which is the region of maximal atmospheric transparency in the infrared and also roughly aligns with the peak TE wavelengths for typical ambient-temperature emitters. For an emitter to cool in daylight, its absorptivity (and hence emissivity) in the visible and near-infrared range also needs to be suppressed, to minimize heating by incident solar radiation [Fig. 5(a, b)].

Early examples of engineered radiative coolers used simple bulk materials, such as pigmented paints (see ref. [117] and references therein for a comprehensive review of those results). More recently, radiative cooling technologies have made use of nanostructured materials, with examples including resonant SiC and $SiO_2$ nanoparticles [118] and a hybrid structure comprising a lossy photonic crystal backed by a thin-film reflector [119].

The nanophotonic approach for radiative cooling has been developed in subsequent works utilizing layered systems [120], metasurfaces [121], photonic crystals [122], $SiO_2$-microsphere-based random materials [123], and hierarchically porous polymer coatings [124]. In particular, simple planar structures [120,125] have enabled cooling below the ambient temperature [Fig. 5(c)], and even to sub-freezing temperature (40 °C below ambient) through a day-night cycle [126]. This approach can also be made dynamic; a recently proposed structure based on $VO_2$ exhibited low emissivity at temperatures below a target temperature, and high emissivity otherwise, mitigating temperature fluctuations [127]. Applications of radiative cooling can range from cooling of dwellings in hot temperatures [122,126] to enhancing vapor condensation[128] to the thermoregulation of spacecraft [129,130].

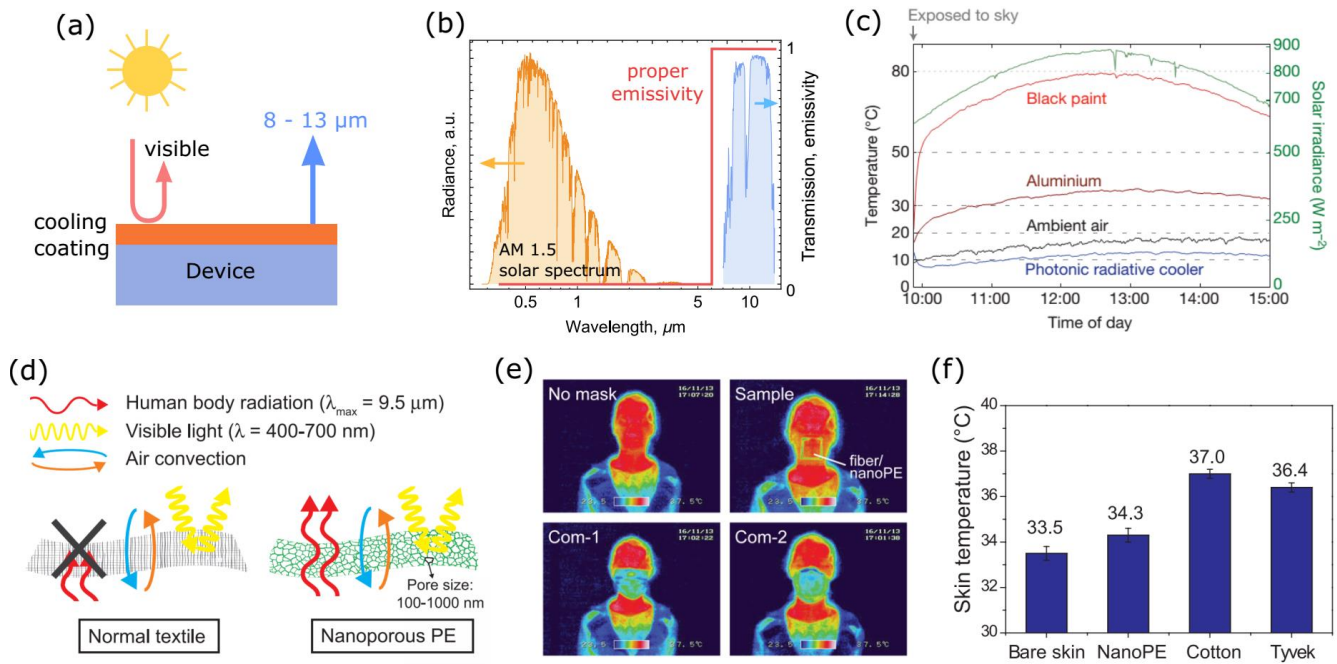

**Figure 5. Radiative cooling.** (a) Schematic of a radiative cooling device: the coating reflects the visible part of the solar spectrum but has high emissivity in the mid-infrared atmospheric-transparency window. (b) The AM 1.5 solar spectrum plotted along with a typical infrared transmission spectrum of the atmosphere (recorded at the Gemini Observatory), which has a transparency window in the 8 – 13 $\mu m$ range, and an appropriate emissivity spectrum required for efficient daytime radiative cooling. (c) Temperature of different thermal emitters throughout on a clear day in California, including a photonic radiative cooler based on a thin-film multilayer that achieves cooling below the ambient temperature. (d) Transmittance of a traditional textile, and a nanoporous polyethylene (nano-PE) material with enhanced infrared transmittance, enabling radiative cooling through the textile. (e) Long-wave infrared images of a human face without a mask, with two commercial masks, and with a nano-PE mask which is semi-transparent to infrared TE. (f) Equilibrium skin temperatures under ambient conditions for the different textile materials. Images reprinted with permissions from: c – ref. [120]; d,f - ref. [131]; e - ref. [132].

Similar concepts can be applied to textiles, enabling efficient thermal management. Traditional textiles are opaque in the infrared, preventing efficient radiation of TE from parts of the human body [Fig. 5(d)]. Recently, fiber-based textiles that enable high infrared transmittance and are opaque in the visible have been presented theoretically [133] and demonstrated experimentally [131]. A similar fiber-based material was incorporated for design of face masks that enable cooling based on infrared TE [132] [Fig. 5(e)]. Such materials can lower skin temperature by several degrees compared to cotton textiles, coming close to the temperature of bare skin [Fig. 5(f)]. Conversely, for cold environments, metallic nanowire-coated [134] and nanoporous metallized [135] textiles have been developed to suppress infrared TE from the body. Finally, a dual-mode invertible textile capable of enhanced cooling or heating has also been recently demonstrated [136]. To achieve this dual-mode operation, one side of the textile was designed to have high infrared emissivity, whereas the other side was designed to have low emissivity, respectively. Heating or cooling is then realized by choosing the right side of the textile facing the outside.

**Thermal camouflage.** TE management with structured materials may also find applications for infrared privacy and camouflage applications. Tuning TE affects the visibility of objects to infrared cameras, e.g., those working in the atmospheric transparency window (8 - 13 $\mu$m). Therefore, the active tuning methods described above can also be used to obscure features from infrared imagers. For example, the negative-differential TE enabled by $VO_2$-based structures [60] can be combined with local temperature control to enable active-camouflage pixels [137]. Alternatively, an all-passive approach using zero-differential TE based on phase-transition materials has been recently demonstrated [81]. Finally, thermal camouflage can also be achieved without emissivity modulation by manipulating surface temperature via transformation thermodynamics [138]. This approach exploits the thermal conductivity in the system, adjusting the profile of the surface temperature to that of the background, thus concealing TE from the object.

## Outlook

In this paper, we reviewed the fundamental properties of thermal emission (TE) from heated bodies, approaches to control its bandwidth and directivity using nanophotonic structures, and applications of TE control in the areas of energy harvesting, lighting, thermal management, and infrared camouflage. We anticipate that TE engineering will continue to enhance technologies where radiative heat transfer can play an important role.

The field of TE engineering has a number of directions that have yet to be fully explored. One intriguing opportunity is the potential to engineer the photon statistics of quasi-coherent TE. Although narrowband TE as discussed in this review exhibits a narrow energy density spectrum and enhanced coherence length, this coherence does not change the quantum statistics of thermal light, which remains super-Poissonian. However, recent theoretical work indicated that statistical behavior of thermal photons may be different in the ultrastrong coupling regime, leading to either a super- or a sub-Poissonian second-order correlation function[139]. Several groups have also discussed possible TE analogies of quantum super-radiant emission [140,141]. Finally, we imagine that there are significant opportunities for TE engineering using various non-reciprocal systems, e.g., based on magneto-optic effects[142] or using temporal modulation[142]. In non-reciprocal structures, detailed balance—the principle that yields Kirchhoff's law—can be violated for certain directions; e.g., the emissivity and reflectivity toward a specific far-field angle can be made to be simultaneously high [142]. This concept has already begun to be explored to improve photovoltaic performance, where TE from the solar cell back into the sky is a loss mechanism [143].

M.A.K. acknowledges financial support from the NSF (ECCS-1750341) and ONR (N00014-16-1-2556). A.K. and A.A. acknowledge support from the AFOSR (MURI grant No. FA9550-17-1-0002), the Simons Foundation and the NSF. D.G.B. acknowledges support from the Knut and Alice Wallenberg Foundation. We acknowledge Susumu Noda for sending data for our figures, and also helpful discussions with Andrej Lenert.

*mkats@wisc.edu

§D.G.B. and Y.X. contributed equally to this manuscript. Correspondence should be addressed to M.K.